\documentclass[11pt]{cernart}
\usepackage{epsfig}
\usepackage{amssymb}
\begin {document}
%
%
\dimen\footins=\textheight
\begin{titlepage}
\docnum{CERN--PH--EP/2007--024}
\date{09 July 2007}
\hbox to \hsize{\hskip123mm\hbox{\small revised 25 October 2007}\hss}
\vspace*{1cm}
\begin{center}
{\LARGE {\bf The Polarised Valence Quark Distribution \\ 
from Semi-inclusive DIS}}
\vspace*{0.5cm}
\normalsize
\end{center}

\author{\large The COMPASS Collaboration}
\vspace{2cm}

\begin{abstract}
The semi-inclusive difference asymmetry
$A^{h^{+}-h^{-}}$ for hadrons of opposite charge has been measured by the
COMPASS experiment at CERN. The data were collected in the years 2002--2004
using a 160~GeV polarised muon beam scattered off a large polarised $^6$LiD target
in the kinematic range $0.006 < x < 0.7$ and $1 < Q^2 <100$~(GeV$/c)^2$.
In leading order QCD (LO) the deuteron asymmetry $A^{h^{+}-h^{-}}$ 
measures the valence quark polarisation and provides an evaluation of 
the first moment of $\Delta u_v + \Delta d_v$ which is found to be
equal to $0.40 \pm 0.07~({\rm stat.}) \pm 0.06~({\rm syst.})$ over 
the measured range of $x$ at $Q^2 = 10$~(GeV$/c)^2$.
When combined with the first moment of $g_1^d$ previously measured on 
the same data, this result favours 
 a non-symmetric polarisation of light quarks 
$\Delta {\overline u} = - \Delta {\overline d}$ 
at a confidence level of two standard deviations,
in contrast to the often 
assumed symmetric scenario 
$\Delta {\overline u} = \Delta {\overline d} = \Delta {\overline s} = \Delta s$.   
\vfill
\submitted{(Submitted to Physics Letters B)}

\end{abstract}
\begin{Authlist}
{\large  The COMPASS Collaboration}\\[\baselineskip]
%
%
M.~Alekseev\Iref{turin_p},
V.Yu.~Alexakhin\Iref{dubna},
Yu.~Alexandrov\Iref{moscowlpi},
G.D.~Alexeev\Iref{dubna},
A.~Amoroso\Iref{turin_u},
A.~Arbuzov\Iref{dubna},
B.~Bade\l ek\Iref{warsaw},
F.~Balestra\Iref{turin_u},
J.~Ball\Iref{saclay},
J.~Barth\Iref{bonnpi}
% ...
\end{Authlist}
%
%
\Instfoot{bielefeld}{Universit\"at Bielefeld, Fakult\"at f\"ur Physik, 33501 Bielefeld, Germany\Aref{e}}
\Instfoot{bochum}{Universit\"at Bochum, Institut f\"ur Experimentalphysik, 44780 Bochum, Germany\Aref{e}}
\Instfoot{bonniskp}{Universit\"at Bonn, Helmholtz-Institut f\"ur  Strahlen- und Kernphysik, 53115 Bonn, Germany\Aref{e}}
\Instfoot{bonnpi}{Universit\"at Bonn, Physikalisches Institut, 53115 Bonn, Germany\Aref{e}}
\Instfoot{brno}{Institute of Scientific Instruments, AS CR, 61264 Brno, Czech Republic\Aref{f}}
\Instfoot{burdwan}{Burdwan University, Burdwan 713104, India\Aref{g}}
\Instfoot{calcutta}{Matrivani Institute of Experimental Research \& Education, Calcutta-700 030, India\Aref{h}}
\Instfoot{dubna}{Joint Institute for Nuclear Research, 141980 Dubna, Moscow region, Russia}
\Instfoot{erlangen}{Universit\"at Erlangen--N\"urnberg, Physikalisches Institut, 91054 Erlangen, Germany\Aref{e}}
\Instfoot{freiburg}{Universit\"at Freiburg, Physikalisches Institut, 79104 Freiburg, Germany\Aref{e}}
\Instfoot{cern}{CERN, 1211 Geneva 23, Switzerland}
\Instfoot{heidelberg}{Universit\"at Heidelberg, Physikalisches Institut,  69120 Heidelberg, Germany\Aref{e}}
\Instfoot{helsinki}{Helsinki University of Technology, Low Temperature Laboratory, 02015 HUT, Finland  and University of Helsinki, Helsinki Institute of  Physics, 00014 Helsinki, Finland}
\Instfoot{liberec}{Technical University in Liberec, 46117 Liberec, Czech Republic\Aref{f}}
\Instfoot{lisbon}{LIP, 1000-149 Lisbon, Portugal\Aref{i}}
\Instfoot{mainz}{Universit\"at Mainz, Institut f\"ur Kernphysik, 55099 Mainz, Germany\Aref{e}}
\Instfoot{miyazaki}{University of Miyazaki, Miyazaki 889-2192, Japan\Aref{j}}
\Instfoot{moscowlpi}{Lebedev Physical Institute, 119991 Moscow, Russia}
\Instfoot{munichlmu}{Ludwig-Maximilians-Universit\"at M\"unchen, Department f\"ur Physik, 80799 Munich, Germany\AAref{e}{k}}
\Instfoot{munichtu}{Technische Universit\"at M\"unchen, Physik Department, 85748 Garching, Germany\AAref{e}{k}}
\Instfoot{nagoya}{Nagoya University, 464 Nagoya, Japan\Aref{j}}
\Instfoot{praguecu}{Charles University, Faculty of Mathematics and Physics, 18000 Prague, Czech Republic\Aref{f}}
\Instfoot{praguectu}{Czech Technical University in Prague, 16636 Prague, Czech Republic\Aref{f}}
\Instfoot{protvino}{State Research Center of the Russian Federation, Institute for High Energy Physics, 142281 Protvino, Russia}
\Instfoot{saclay}{CEA DAPNIA/SPhN Saclay, 91191 Gif-sur-Yvette, France}
\Instfoot{telaviv}{Tel Aviv University, School of Physics and Astronomy, 69978 Tel Aviv, Israel\Aref{l}}
\Instfoot{triest_i}{Trieste Section of INFN, 34127 Trieste, Italy}
\Instfoot{triest}{University of Trieste, Department of Physics and Trieste Section of INFN, 34127 Trieste, Italy}
\Instfoot{triestictp}{Abdus Salam ICTP and Trieste Section of INFN, 34127 Trieste, Italy}
\Instfoot{turin_u}{University of Turin, Department of Physics and Torino Section of INFN, 10125 Turin, Italy}
\Instfoot{turin_i}{Torino Section of INFN, 10125 Turin, Italy}
\Instfoot{turin_p}{University of Eastern Piedmont, 1500 Alessandria,  and Torino Section of INFN, 10125 Turin, Italy}
\Instfoot{warsaw}{So{\l}tan Institute for Nuclear Studies and Warsaw University, 00-681 Warsaw, Poland\Aref{m} }
\Instfoot{warsawtu}{Warsaw University of Technology, Institute of Radioelectronics, 00-665 Warsaw, Poland\Aref{n} }
\Instfoot{yamagata}{Yamagata University, Yamagata, 992-8510 Japan\Aref{j} }
%
%
\Anotfoot{+}{Deceased}
\Anotfoot{a}{Also at IST, Universidade T\'ecnica de Lisboa, Lisbon, Portugal}
\Anotfoot{b}{On leave of absence from JINR Dubna}
\Anotfoot{c}{Also at Chubu University, Kasugai, Aichi, 487-8501 Japan}
\Anotfoot{d}{Also at also at GSI mbH, Planckstr.\ 1, D-64291 Darmstadt, Germany}
\Anotfoot{e}{Supported by the German Bundesministerium f\"ur Bildung und Forschung}
\Anotfoot{f}{Suppported by Czech Republic MEYS grants ME492 and LA242}
\Anotfoot{g}{Supported by DST-FIST II grants, Govt. of India}
\Anotfoot{h}{Supported by  the Shailabala Biswas Education Trust}
\Anotfoot{i}{Supported by the Portuguese FCT - Funda\c{c}\~ao para a Ci\^encia e Tecnologia grants POCTI/FNU/49501/2002 and POCTI/FNU/50192/2003}
\Anotfoot{j}{Supported by the Ministry of Education, Culture, Sports, Science and Technology, Japan, Grant-in-Aid for Specially Promoted Research No.\ 18002006; Daikou Foundation and Yamada Foundation}
\Anotfoot{k}{Supported by the DFG cluster of excellence `Origin and Structure of the Universe' (www.universe-cluster.de)}
\Anotfoot{l}{Supported by the Israel Science Foundation, founded by the Israel Academy of Sciences and Humanities}
\Anotfoot{m}{Supported by KBN grant nr 621/E-78/SPUB-M/CERN/P-03/DZ 298 2000, nr 621/E-78/SPB/CERN/P-03/DWM 576/2003-2006, and by MNII reasearch funds for 2005--2007}
\Anotfoot{n}{Supported by KBN grant nr 134/E-365/SPUB-M/CERN/P-03/DZ299/2000}

\vfill

\newpage
~
\end{titlepage}

\noindent
The COMPASS experiment at CERN has recently published an evaluation of the
deuteron spin-dependent structure function $g_1^d(x)$ in the DIS region,
based on measurements of the spin asymmetries observed in the scattering
of 160~GeV longitudinally polarised 
muons on a longitudinally polarised $^6$LiD target \cite{g1d}. These measurements
provide an accurate evaluation of the first moment of $g_1$ for the
average nucleon  $N$ in an isoscalar target $g_1^N = (g_1^p + g_1^n)/2$
\begin{equation}
  \Gamma_1^N(Q^2\!=\! 10~({\rm GeV}/c)^2) = 
  \int_0^1 g_1^N(x,Q^2\!=\! 10~({\rm GeV}/c)^2) {\rm d}x = 
  0.051 \pm 0.003~{\rm (stat.)}\pm 0.006~{\rm (syst.)}
\label{gam}
\end{equation}
from which the  first moment of the strange quark distribution    
can be extracted if the value of the octet matrix element 
($a_8 = 3 F - D$) is taken from semi-leptonic hyperon decays.%
\footnote {At the precision of the experiment the value of $\Gamma_1^N$
is unchanged when the  evolution of the measured values $g_1(x_i,Q^2_i)$
to a common $Q^2$ 
is done at LO or at NLO in QCD.}
At LO in QCD the strange quark polarisation is given by

\begin{equation}
  \Delta s + \Delta {\overline s} = 
  3 \Gamma_1^N - \frac{5}{12} a_8 =
  -0.09 \pm 0.01~(\rm{stat.}) \pm 0.02~(\rm{syst.})
\label{def_Ds}
\end{equation}
at $Q^2 = 10$~(GeV$/c)^2$.

Since quarks and antiquarks of the same flavour 
equally contribute to $g_1$, inclusive data 
do not allow to separate valence and sea contributions to the nucleon spin.
We present here additional information on the contribution of the nucleon 
constituents to its spin based on semi-inclusive spin asymmetries measured on 
the same data as those used in Ref.~\cite{g1d}.

The semi-inclusive spin asymmetries for positive and negative hadrons 
$h^+$ and $h^-$ are defined by
\begin{eqnarray}
  A^{h^+} = \frac{\sigma_{\uparrow\downarrow}^{h+}-\sigma_{\uparrow\uparrow}^{h+}}{\sigma_{\uparrow\downarrow}^{h+}+\sigma_{\uparrow\uparrow}^{h+}},
~~~~~~~~~~~~
A^{h^-} = \frac{\sigma_{\uparrow\downarrow}^{h-}-\sigma_{\uparrow\uparrow}^{h-}}{\sigma_{\uparrow\downarrow}^{h-}+\sigma_{\uparrow\uparrow}^{h-}},
\label{def_hadron_asym}
\end{eqnarray}
where the arrows indicate the relative beam and target spin orientations.

The data used in the present analysis were collected by the COMPASS collaboration
at CERN during the years 2002--2004. The event selection requires a 
reconstructed interaction vertex defined by the incoming and scattered muons
and located inside one of the two target cells \cite{spectro}. The energy
of the beam muon is required to be in the interval $140~{\rm GeV} < E_{\mu} < 180$~GeV
and its extrapolated trajectory is required to cross entirely the two cells
in order to equalise the fluxes seen by each of them. DIS events are selected
by cuts on the photon virtuality ($Q^2 > 1$~(GeV$/c)^2$) and on the fractional 
energy of the virtual photon ($0.1 < y < 0.9$).
Final state muons are identified by signals collected behind the hadron absorbers.
 The hadrons used in the
analysis are required to originate from the interaction vertex and 
to be produced in the current fragmentation region. The latter requirement is
satisfied by selecting hadrons with
fractional energy $z > 0.2$.  
In addition an upper limit $z < 0.85$ is
imposed in order to suppress hadrons from exclusive diffractive processes
and to avoid contamination from 
 muons close to the beam axis which escape identification by the muon filters.
The hadron identification
provided by the RICH detector is not used in the present analysis.
 The resulting sample contains 30 and 25 million
of positive and negative hadrons, respectively.

\begin{figure}
  \includegraphics[width=0.49\textwidth,clip]{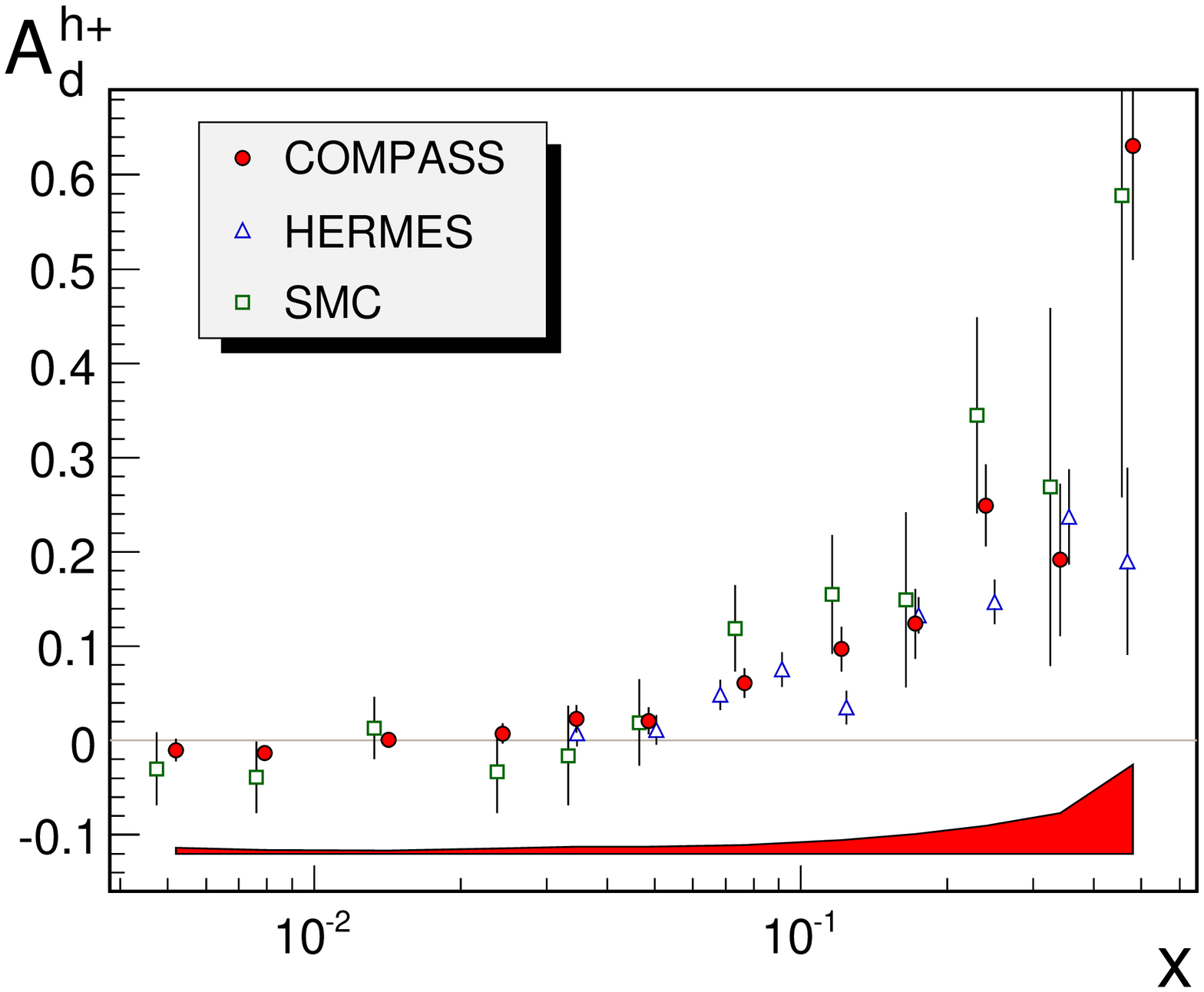}
  \hfill
  \includegraphics[width=0.49\textwidth,clip]{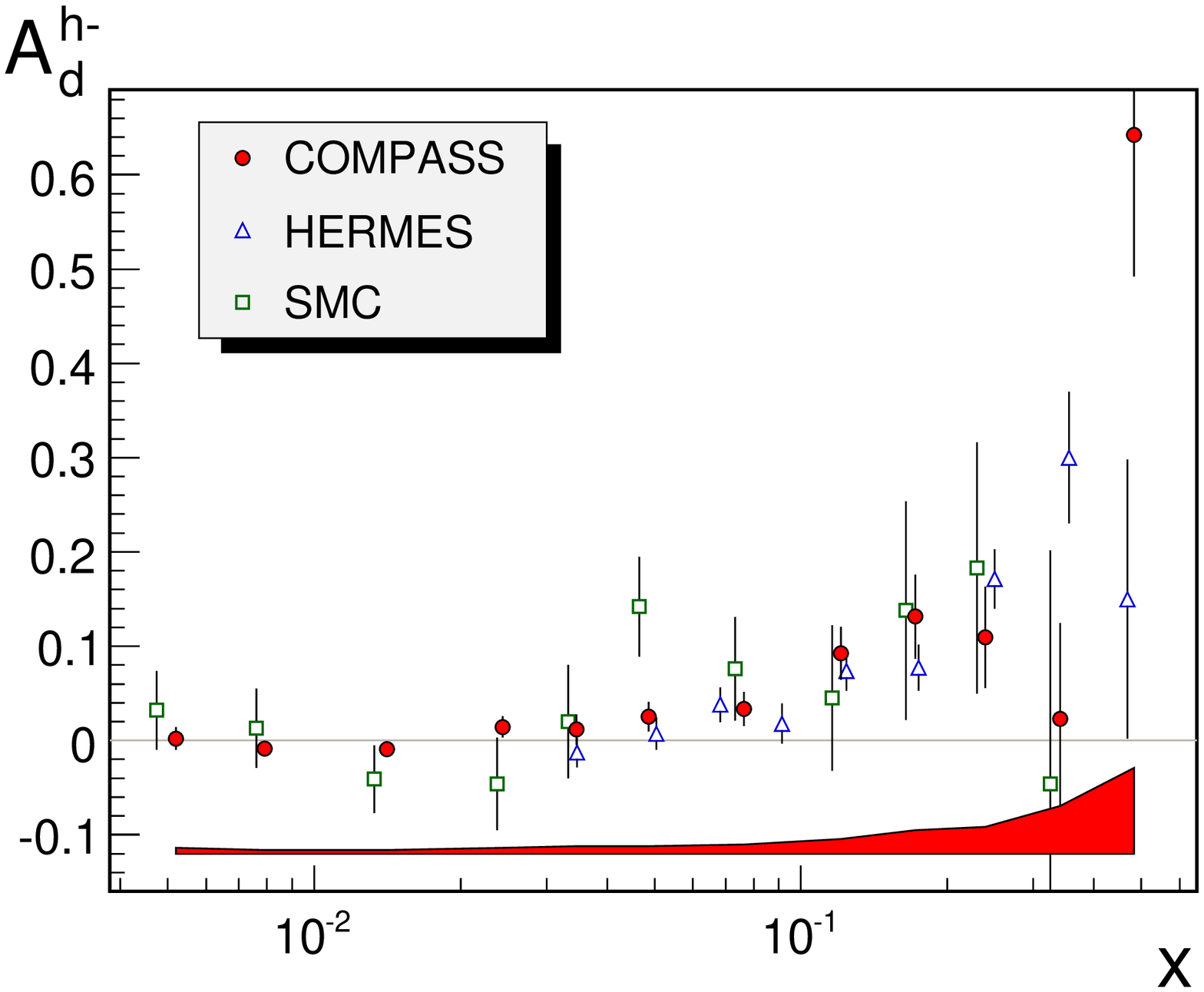}
  \caption{Hadron asymmetries $A_{d}^{h+}$ (left) and 
    $A_{d}^{h-}$ (right) measured  by COMPASS,
    SMC \cite{SMC98} and HERMES \cite{HERMES} experiments.
    The bands at the bottom of the figures show the systematic 
    errors of the COMPASS measurements.
  }
  \label{fig:Ahpm}
\end{figure}

The target spins are reversed at regular intervals of 8 hours during the data
taking.
The spin asymmetries are obtained from the numbers of hadrons
collected from each target cell during consecutive periods 
before and after reversal of the target spins,
following the same procedure as for inclusive asymmetries \cite{a1d}.
They are listed in Table~\ref{tab:asym} and also shown in Fig.~\ref{fig:Ahpm}
as a function of $x$, in comparison to the SMC
\cite{SMC96,SMC98} and HERMES \cite{HERMES} results. 
The consistency of the results from the three experiments 
illustrates the weak $Q^2$ dependence of the semi-inclusive asymmetries. 
The COMPASS results show a large gain in statistical precision with respect
to SMC, especially  in the low $x$ region ($x < 0.04$), while at larger $x$ the COMPASS
errors are comparable to those of HERMES.
The systematic errors, shown by the bands at the bottom of the figure, result
from different sources. The uncertainty on the various factors entering in
the asymmetry calculation (beam and target polarisation, depolarisation factor
and dilution factor) leads to a relative error of 8\% 
on the asymmetry when combined in quadrature.
The uncertainty due to radiative corrections is smaller than in 
the inclusive case due to the selection of hadronic
events and does not exceed $10^{-3}$ in any $x$ bin.
The presence of possible false asymmetries due to time-dependent 
apparatus effects has been studied in the same way as
for the inclusive asymmetries: the data sample has been divided into a large number
of subsamples, each of them collected in a small time interval. The observed 
dispersion of the asymmetries obtained for these subsamples has been found
compatible with the value expected from their statistical error. This allows 
to set an upper limit for this type of  false asymmetries at about half of the
statistical error.  Asymmetries, obtained with different settings of the 
microwave frequency used for dynamic nuclear polarisation of the target, have
also been compared and did not reveal any systematic difference.

 Under the common assumption that hadrons in the current fragmentation region are 
produced by independent quark fragmentation, the semi-inclusive asymmetries $A^{h^+}$,$A^{h^-}$
can be written in LO approximation as
\begin{eqnarray}
 A^h(x,z,Q^2) = \frac{\sum_q e^2_q \Delta q(x,Q^2) D_q^h(z,Q^2)}
  {\sum_q e^2_q  q(x,Q^2) D_q^h(z,Q^2)}
\label{indep_frag}
\end{eqnarray}
where $\Delta q(x,Q^2)$ and $q(x,Q^2)$ are the polarised and unpolarised parton
distribution functions (pdf's) and $D_q^h(z,Q^2)$ the fragmentation function of
a parton $q$ into a hadron $h$. The above formula does not account for 
the full complexity of the hadronisation process as described, for instance in the
Lund string fragmentation model \cite{jetset}, and its validity in low energy
fixed target experiments has been questioned \cite{aram_1}. It has nevertheless
been shown to hold as a good approximation at the energy of COMPASS 
\cite{aram_2}.

In addition to purely experimental effects such as ($x,z$) correlations
in the spectrometer acceptance,
a $z$ dependence of the semi-inclusive asymmetries $A^{h+},A^{h-}$ 
could  reveal a  
breakdown of the independent fragmentation formula (Eq.~(\ref{indep_frag})).
In order to check  the possible presence of this effect, we have re-evaluated
the asymmetries for each $x$ bin subdivided into three intervals of $z$. 
Within their statistical precision the obtained values
 do not indicate any systematic
$z$ dependence.

\begin{table}[]
\begin{center}
\begin{tabular}{|l|c|r|r|r|}
\hline  \hline
\multicolumn{1}{|c|}{$\langle x \rangle$} & $\langle Q^2 \rangle $  & \multicolumn{1}{c|}{$A_d^{h+}$} & \multicolumn{1}{c|}{$A_d^{h-}$} & \multicolumn{1}{c|}{$A_d^{h^+-h^-}$} \\
                     &  (GeV$/c)^2$      &               &      &       \\
\hline
0.0052  &  1.17 &$ -0.010 \pm 0.012 \pm 0.006$ & $ 0.002 \pm 0.012 \pm 0.006$ &   \multicolumn{1}{|c|}{--} \\
0.0079  &  1.45 &$ -0.013 \pm 0.008 \pm 0.004$ & $-0.008 \pm 0.008 \pm 0.004$ & $-0.081 \pm 0.138 \pm 0.070$ \\
0.0141  &  2.06 &$  0.000 \pm 0.007 \pm 0.003$ & $-0.009 \pm 0.007 \pm 0.004$ & $ 0.070 \pm 0.067 \pm 0.034$ \\
0.0244  &  2.99 &$  0.007 \pm 0.011 \pm 0.005$ & $ 0.014 \pm 0.012 \pm 0.006$ & $-0.027 \pm 0.077 \pm 0.039$ \\
0.0346  &  4.03 &$  0.023 \pm 0.015 \pm 0.008$ & $ 0.012 \pm 0.016 \pm 0.008$ & $ 0.070 \pm 0.090 \pm 0.045$ \\
0.0486  &  5.56 &$  0.021 \pm 0.014 \pm 0.007$ & $ 0.025 \pm 0.016 \pm 0.008$ & $ 0.006 \pm 0.076 \pm 0.038$ \\
0.0764  &  8.29 &$  0.061 \pm 0.016 \pm 0.009$ & $ 0.033 \pm 0.018 \pm 0.009$ & $ 0.138 \pm 0.070 \pm 0.037$ \\
0.121   & 12.6  &$  0.097 \pm 0.024 \pm 0.014$ & $ 0.092 \pm 0.028 \pm 0.016$ & $ 0.107 \pm 0.087 \pm 0.044$ \\
0.172   & 17.7  &$  0.124 \pm 0.037 \pm 0.021$ & $ 0.132 \pm 0.045 \pm 0.025$ & $ 0.109 \pm 0.121 \pm 0.061$ \\
0.239   & 25.3  &$  0.249 \pm 0.044 \pm 0.029$ & $ 0.109 \pm 0.054 \pm 0.028$ & $ 0.478 \pm 0.130 \pm 0.075$ \\
0.341   & 42.6  &$  0.192 \pm 0.081 \pm 0.043$ & $ 0.023 \pm 0.101 \pm 0.051$ & $ 0.429 \pm 0.217 \pm 0.114$ \\
0.482   & 60.2  &$  0.630 \pm 0.121 \pm 0.078$ & $ 0.643 \pm 0.150 \pm 0.091$ & $ 0.616 \pm 0.291 \pm 0.186$ \\
\hline  \hline
\end{tabular}
\caption{
  Values of $A_d^{h^+}$, $A_d^{h^-}$ and $A_d^{h^+-h^-}$ with their 
  statistical and systematical errors as a function of $x$ with 
  the corresponding average value of $Q^2$.}
\label{tab:asym}
\end{center}
\end{table}

In the present analysis we use the  ``difference asymmetry'' 
which is defined as the spin asymmetry for the difference of the cross sections 
for positive and negative hadrons
\begin{eqnarray}
A^{h^+ - h^-} = \frac{(\sigma_{\uparrow\downarrow}^{h+}-\sigma_{\uparrow\downarrow}^{h-})-(\sigma_{\uparrow\uparrow}^{h+}-\sigma_{\uparrow\uparrow}^{h-})}{(\sigma_{\uparrow\downarrow}^{h+}-\sigma_{\uparrow\downarrow}^{h-})+(\sigma_{\uparrow\uparrow}^{h+}-\sigma_{\uparrow\uparrow}^{h-})} .
\label{def_diff_asym}
\end{eqnarray}
The difference asymmetry approach for the extraction of helicity distributions,
 introduced 
in Ref.~\cite{Frankfurt}, 
has been used in the SMC analysis \cite{SMC96}  and been further discussed in \cite{Christova,SSI}. 
In LO QCD and under the assumption of isospin and charge conjugation symmetries,
 the fragmentation functions  $D_q^h$ cancel out from $A^{\pi^+ - \pi^-}$.
In addition, in the case of  an isoscalar target
and assuming $\Delta s = \Delta {\overline s}$,
the difference asymmetries  for pions and kaons are both equal to the
valence quark polarisation 
\begin{equation}
  A_{N}^{\pi^+ - \pi^-} = A_N^{K^+ - K^-} = \frac{\Delta u_v + \Delta d_v}{u_v + d_v} \,,
\label{diff_deuteron}
\end{equation}
where we introduce the valence quark distributions $q_v=q-\bar{q}$.
Since kaons contribute to the asymmetry in the same way as pions, 
 their  identification is  not needed, allowing to
reduce the statistical errors.
 The difference asymmetry for
(anti)protons $A_N^{p - {\overline p}}$
has also the same value but under  more restrictive assumptions and is more likely
to be affected by target remnants. Since protons and antiprotons account
only for about 10\% of the selected hadron sample, the relation 
\begin{equation} 
  A_{N}^{h^+ - h^-} \approx \frac{\Delta u_v + \Delta d_v}{u_v + d_v}
\end{equation}
is expected to hold as a good approximation in the present analysis.
Monte Carlo studies using POLDIS \cite{poldis} and Lund string hadronisation
show that the asymmetries  $A_N^{p - {\overline p}}(x)$ 
closely follow the trend of  $A_{N}^{\pi^+ - \pi^-}(x)$ with a difference
never exceeding 0.02. In addition the semi-inclusive asymmetries $A_{N}^{h^+ - h^-}$
are found to be very close to the expected values $(\Delta u_v + \Delta d_v)/(u_v + d_v)$
 defined by the input parametrisations
in the Monte Carlo simulation with the largest difference ($\le 0.05$)
appearing in the two highest intervals of $x$.   

At higher order in QCD the difference asymmetries
still determine the valence quark polarisation without any assumption
on the sea and gluon densities \cite{Christova}. Fragmentation functions
no longer cancel out 
but their effect is expected to be small \cite{SSI}.

\begin{figure}[h]
  \includegraphics[width=0.49\textwidth,clip]{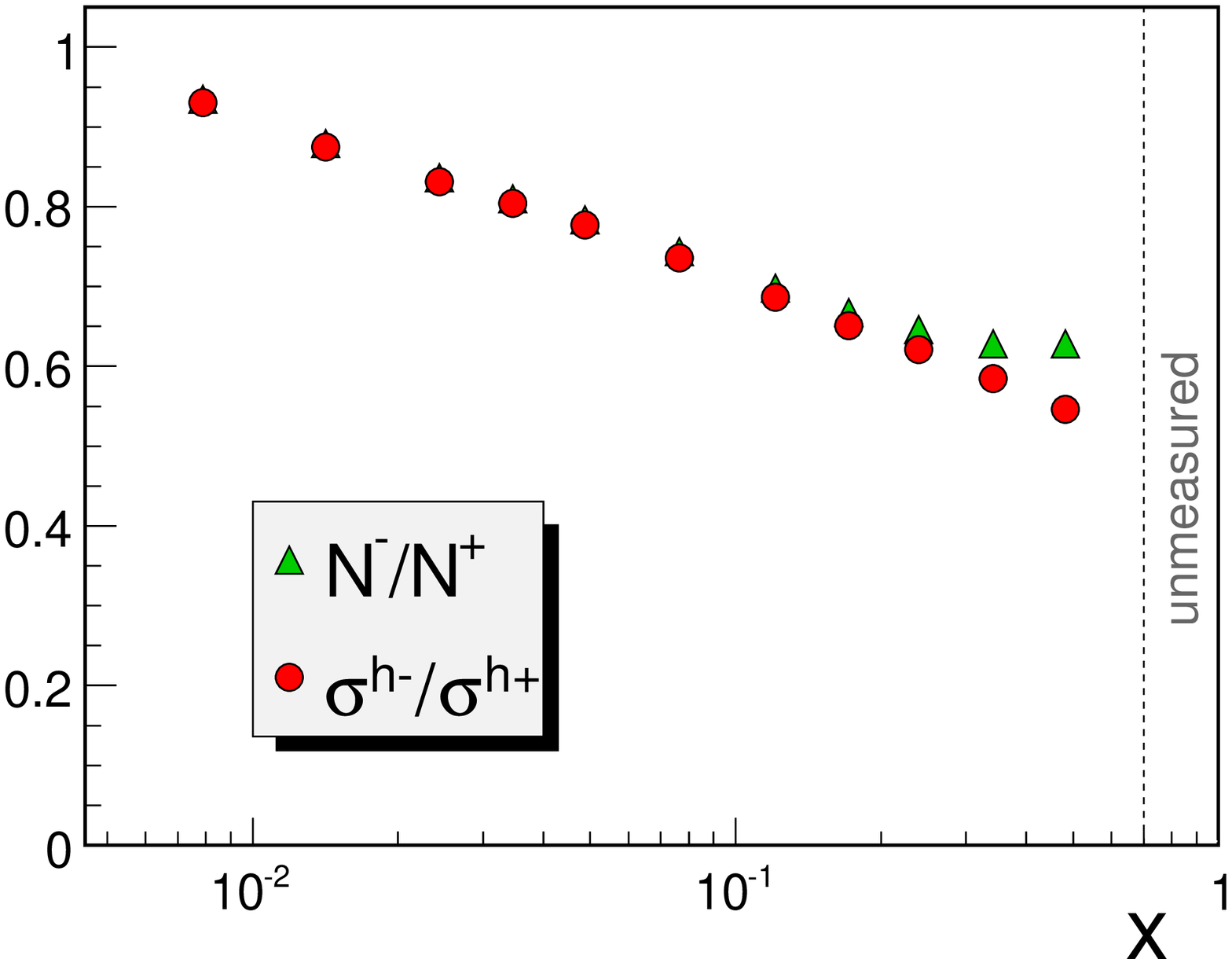}
  \hfill
  \includegraphics[width=0.49\textwidth,clip]{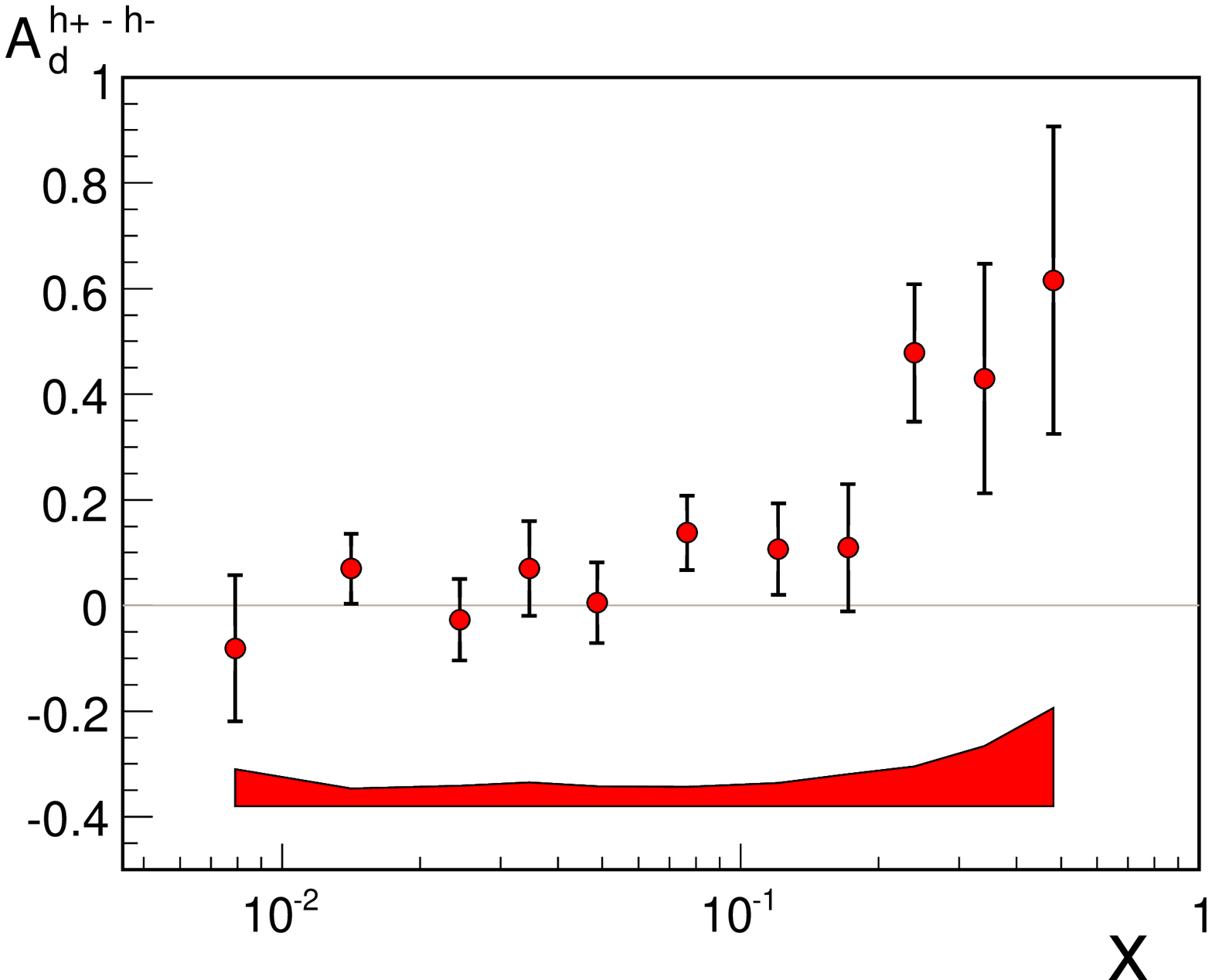}
  \caption{Left: 
    The ratio $\sigma^{h^{-}}/\sigma^{h^{+}}$ before (triangles) and
    after acceptance corrections (circles). 
    Right: The difference asymmetry, $A_d^{h^+-h^-}$, for unidentified
    hadrons of opposite charges, as a function of $x$ at the $Q^2$ of 
    each measured point.
  }
  \label{fig:Ratio_N_release}
\end{figure}

The relation between the difference asymmetries of Eq.~(\ref{def_diff_asym}) and 
the single hadron asymmetries of  Eq.~(\ref{def_hadron_asym}) is
\begin{eqnarray}
A^{h^+-h^-} = \frac{1}{1-r} ( A^{h^+} - r A^{h^-}) \,, & {\rm with}&
r=\frac{\sigma_{\uparrow\downarrow}^{h-}+\sigma_{\uparrow\uparrow}^{h-}}
{\sigma_{\uparrow\downarrow}^{h+}+\sigma_{\uparrow\uparrow}^{h+}} =
\frac{\sigma^{h-}}{\sigma^{h+}} .
\label{rel_diff_asym}
\end{eqnarray}
The ratio of cross sections for negative and positive hadrons $r$
depends on the event kinematics and 
is obtained as the product of the corresponding ratio of the number of observed 
hadrons $N^-/N^+$  by the ratio of the geometrical acceptances $a^+/a^-$
\begin{eqnarray}
r  = \frac{\sigma^{h-}}{\sigma^{h+}}  = 
\frac{N^-}{N^+} \cdot \frac{a^+}{a^-}. 
\label{ratio_indirect}
\end{eqnarray}
 The  ratio of the number 
of negative to positive hadrons (Fig.~\ref{fig:Ratio_N_release}, left) 
 decreases with increasing $x$.
This ratio is subject to acceptance corrections because positive and negative
hadrons, produced at the same angle, cross different regions of the spectrometer. 
To this end LEPTO generated Monte Carlo events have been 
processed through the program simulating the COMPASS spectrometer performance
\cite{spectro} and reconstructed in the same way as the data.
The acceptances $a^+$ and $a^-$
are indeed found to be different. The ratio $a^-/a^+$, which is about 1.0 at low $x$,
increases for $x > 0.1$  reaching $\sim$$1.12$ in the highest $x$ bin.
Bin migration was found to be negligible.
The corrected cross section ratio $\sigma^{h^{-}}/\sigma^{h^{+}}$ is also shown in
Fig.~\ref{fig:Ratio_N_release}.

The resulting values of the difference asymmetry $A_d^{h^+-h^-}$  
as a function of $x$ are shown in Fig.~\ref{fig:Ratio_N_release} (right) 
and listed with their statistical and systematic
errors in Table~\ref{tab:asym}. The statistical correlation between $A_d^{h^+}$ and 
$A_d^{h^-}$ which is approximately 0.20 over the measured range of $x$,  is taken
into account in the evaluation of the error of  $A_d^{h^+-h^-}$. 
 As can be seen from  Eq.~(\ref{rel_diff_asym}), a
singularity appears when the cross section ratio becomes close to one, leading
to infinite statistical errors. For this reason, we discard the lowest $x$ bin
used in the inclusive $g_1$ analysis \cite{g1d} 
and take $x = 0.006$  as lower limit for the 
present analysis. The increase of  $A_d^{h^+-h^-}$
for $x > 0.1$ illustrates the increasing polarisation of valence quarks 
carrying a larger fraction of the nucleon momentum.

Target remnants may  affect current quark fragmentation
at low values of the total hadronic energy $W$. 
In order to check the possible presence of such effects in our data,
 we have  re-evaluated
the difference asymmetries  $A_d^{h^+-h^-}(x)$
with the cut $W \ge 7$~GeV.  
The comparison of the obtained values 
with those quoted in Table~1 shows that the $W$ cut only affects
the two highest intervals of $x$ where $A_d^{h^+-h^-}(x)$ is reduced by about 0.3 $\sigma_{\rm stat}$.
It will be shown below that these two values can 
be replaced by  more accurate estimations so that the observed $W$ dependence
does not affect   any further result.

\begin{figure}[tbp]
  \includegraphics[width=0.49\textwidth,clip]{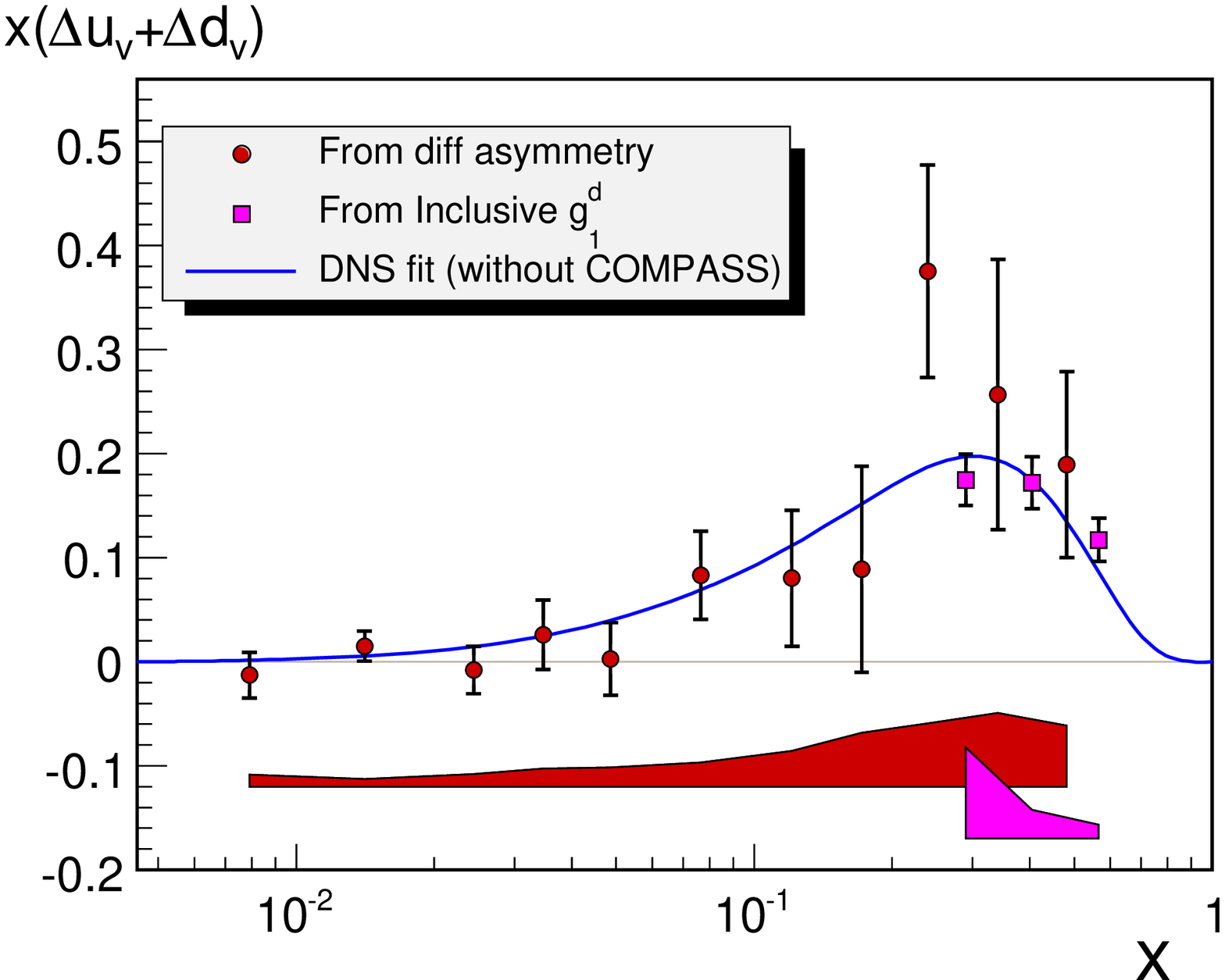}
  \hfill
  \includegraphics[width=0.49\textwidth,clip]{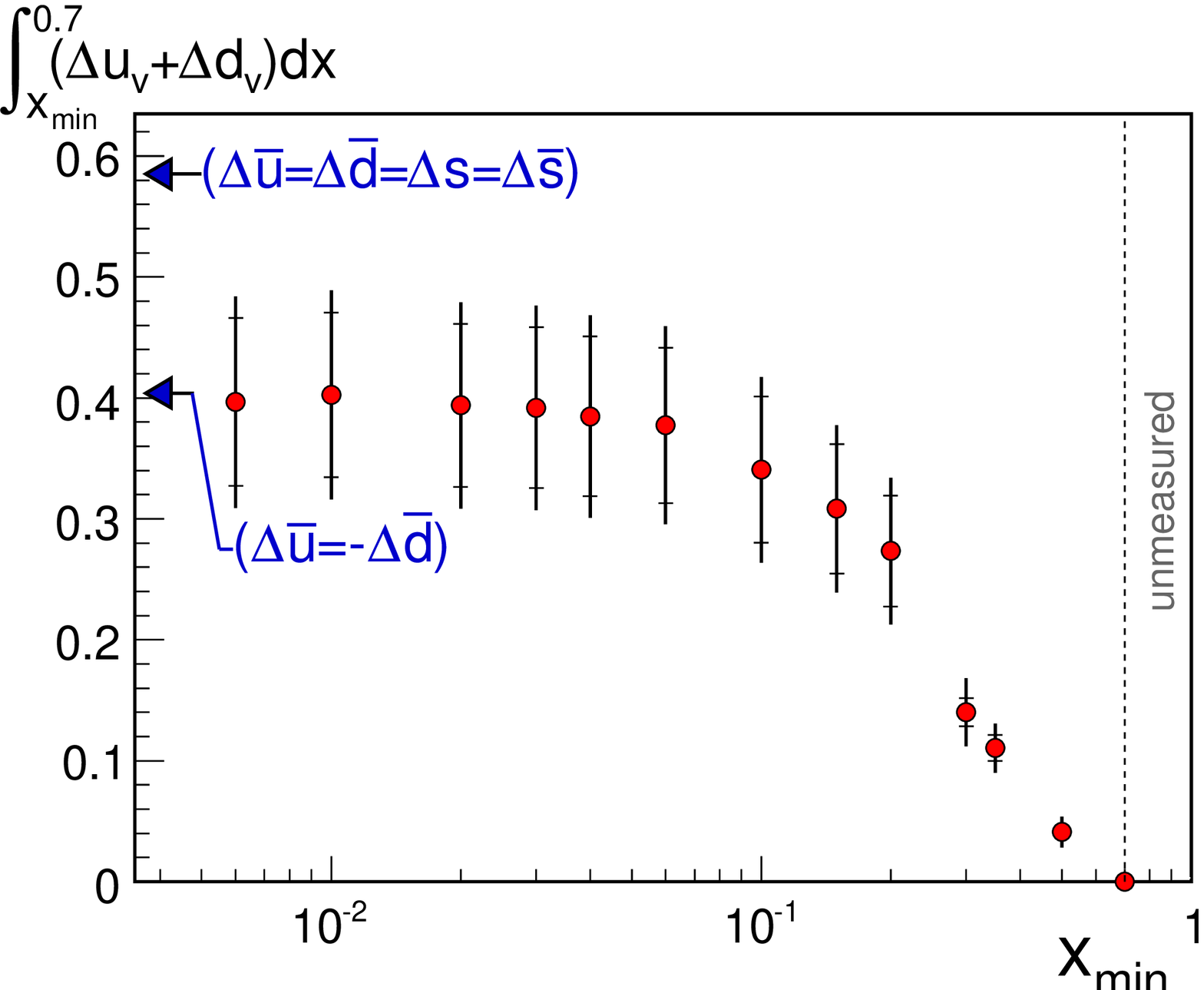}
  \caption{Left:
    Polarised valence quark distribution $x(\Delta u_v(x)+\Delta d_v(x))$
    evolved to $Q^2 = 10$~(GeV$/c)^2$ according to the DNS fit at LO \cite{dns} (line).
    Three additional points at high $x$ are obtained from $g_1^d$ \cite{g1d}.
    The two shaded bands show the systematic errors for the two sets of values.
    Right:
    The integral of $\Delta u_v(x)$$+$$\Delta d_v(x)$ over the range
    $0.006<x<0.7$ as the function of the low $x$ limit, 
    evaluated at $Q^2=10$~(GeV$/c)^2$. 
  }
  \label{fig:Adiff_11_release}
\end{figure}

The polarised valence quark distribution   $\Delta u_v + \Delta d_v$ is
obtained by multiplying $A_d^{h^+-h^-}$ by the unpolarised valence distribution of
MRST04 at LO \cite{mrst}.
Here two corrections are applied, one accounting for the fact that
although $R(x,Q^2) = 0$ at LO, the unpolarised pdf's
originate from $F_2$'s in which $R = \sigma_L/\sigma_T$ was different from zero \cite{abe},
the other one accounting for deuteron D-state contribution ($\omega_D = 0.05\pm0.01$ \cite{mch}):
\begin{equation}
\Delta u_v + \Delta d_v = 
\frac{(u_v + d_v)_{\rm MRST}}{(1 + R) (1 - 1.5 \omega_D)} A_d^{h^+-h^-}.
\label{eq:SIDIS_Apm}
\end{equation}
The LO parameterisation of the DNS fit \cite{dns} has been used to evolve all 
values of $\Delta u_v + \Delta d_v$
to a common $Q^2$ fixed at $Q^2_0 = 10$~(GeV$/c)^2$
 assuming that the difference between $\Delta u_v + \Delta d_v$ at the
current $Q^2$ and  at $Q^2_0$ is the same for the data as for the fit \cite{g1d}.
The  DNS analysis includes all DIS $g_1$ data prior to COMPASS,
the partial COMPASS data on $g_1$ from Ref.~\cite{a1d} as well as
the SIDIS data from SMC \cite{SMC98} and HERMES \cite{HERMES}.
Two parameterisations  of polarised pdf's
are provided at LO, corresponding to two
different choices of fragmentation functions, KRE \cite{kretzer} and KKP \cite{kkp}.
We have checked that the $x$ dependence of the ratio $\sigma^{h^{-}}/\sigma^{h^{+}}$
(Fig.~\ref{fig:Ratio_N_release}) 
is fairly well reproduced by the LO 
MRST04 pdf's  and the KKP  
fragmentation functions  whereas the  KRE parameterisation 
leads to a  much weaker $x$ dependence.
For this reason we choose the fit with the KKP parameterisation.
 The largest corrections to $x[\Delta u_v(x) + \Delta d_v(x)]$ are at
large $x$ and $Q^2$ and do not exceed 0.03. The use of different fits (NLO fit of
Ref.~\cite{dns} or \cite{aac}) leads practically to the same results. 
The resulting values are shown in Fig.~\ref{fig:Adiff_11_release} (left).
The DNS fit, also shown in the figure, is basically defined
by the SMC and HERMES semi-inclusive asymmetries.   
Its  good agreement with the COMPASS values ($\chi^2 = 7.7$ for 11 data points)
illustrates the consistency between the three experiments.

The sea contribution to the unpolarised structure function $F_2$
decreases rapidly with increasing $x$ and
becomes smaller than 0.1 for $x > 0.3$. 
Due to the positivity conditions 
$|\Delta q| \le q$ and $|\Delta {\overline q}| \le {\overline q}$, the
polarised sea contribution to the nucleon spin also becomes negligible in
this region. 
In view of this, the evaluation of the valence spin 
distribution of Eq.~(\ref{eq:SIDIS_Apm}) can be replaced by a more
accurate one obtained from inclusive interactions. Indeed at LO one obtains
\begin{equation}
  \Delta u_v + \Delta d_v ~=~ \frac{36}{5} \frac{g_1^d}{(1 - 1.5 \omega_D)}
  -
  \left[ 2(\Delta\bar{u}+\Delta\bar{d})+\frac{2}{5}(\Delta s +\Delta\bar{s}) \right].
  \label{eq:DIS_g1d}
\end{equation}
The values obtained by taking only the first term on the r.h.s.\ for $x > 0.3$
are also shown in Fig.~\ref{fig:Adiff_11_release}. They  agree very well
with the DNS curve, which is based on previous experiments where the same procedure
had been applied \cite{SMC98,HERMES}. 
 The upper limit of the neglected sea quark contribution, derived from
the saturation of the positivity constraint
 $|\Delta q| \le q$ is included in the systematic error.

The first moment of the polarised valence distribution, truncated to  the measured
range of $x$,
\begin{equation}
\Gamma_v(x_{\rm min}) = \int_{x_{\rm min}}^{0.7} \left[\Delta u_v(x) + \Delta d_v(x)\right] {\rm d}x,
\label{eq:momment}
\end{equation}
derived from the difference asymmetry for $x < 0.3$ and from $g_1^d$
for $0.3 < x< 0.7$,
is shown in Fig.~\ref{fig:Adiff_11_release} (right). Practically no
dependence on the lower limit is observed for $x_{\rm min} < 0.03$.
We obtain for the full measured range of $x$
\begin{equation}
\Gamma_v(0.006 < x < 0.7) = 0.40 \pm 0.07~{\rm (stat.)} \pm 0.06~{\rm (syst.)}
\end{equation}
at $Q^2 = 10$~(GeV$/c)^2$, with contributions of $0.26 \pm 0.07$ and 
$0.14 \pm 0.01$ for $x < 0.3$ and $x > 0.3$, respectively.
The uncertainty due to the unpolarised valence quark distributions ($\approx 0.04$)
has been estimated by comparing different LO parametrisations and been included
in the systematic error.
It should be noted that removing the factor
$(1 + R)$ in Eq.~(10) would increase the value of 
$\Gamma_v$ to $0.42 \pm 0.08 \pm 0.06$. 
Our value of $\Gamma_v$ confirms the HERMES result obtained at 
$Q^2 = 2.5$~(GeV$/c)^2$ over a smaller range of $x$ and is also consistent 
with the  SMC result which has three times larger errors (Table~\ref{tab:FirstMom}). 
The factor $(1+R)$ was also used in the analyses of the previous experiments.

\begin{table}
\begin{tabular}{|l||c|c||c|c|c|c|}
\hline
 & $x$-range & $Q^2$ & \multicolumn{2}{c|}{$\Delta u_v + \Delta d_v$} & \multicolumn{2}{c|}{$\Delta\bar{u} + \Delta\bar{d}$} \\
\cline{4-7}
   &  & $\!\!$(GeV$/c)^2\!\!$       & Exp.Value  & DNS   &    Exp.Value       & DNS \\
\hline
\hline
 SMC & $0.003-0.7$ & 10 & $0.26\pm0.21\pm0.11$ & 0.386 & ~~$0.02\pm0.08\pm0.06$ & $-0.009$ \\
\hline
 HERMES & $0.023-0.6$ & 2.5 & $0.43\pm0.07\pm0.06$ & 0.363 & $-0.06\pm0.04\pm0.03$ & $-0.005$ \\
\hline
\hline
         & $0.006-0.7$ &     & $0.40\pm0.07\pm0.06$  & 0.385 &       --             & $-0.007$ \\
 \raisebox{1.5ex}[-1.5ex]{COMPASS}&       $0-1$ & \raisebox{1.5ex}[-1.5ex]{10} & $0.41\pm0.07\pm0.06$  &  --   & ~$0.0\pm0.04\pm0.03$ &    --    \\
\hline
\end{tabular}
\caption{Estimates of the  first moments $\Delta u_v + \Delta d_v$
  and $\Delta\bar{u} + \Delta\bar{d}$ from the SMC \cite{SMC98}, HERMES \cite{HERMES},
  COMPASS data and also from the DNS fit at LO \cite{dns} truncated to the
  range of each experiment (lines 1--3). 
  The SMC results were obtained with the assumption of a $SU(3)_f$ symmetric sea:
  $\Delta\bar{u}=\Delta\bar{d}=\Delta\bar{s}$.
The last line shows the COMPASS results for the full range of $x$.
}
\label{tab:FirstMom}
\end{table}

The difference between our measured value of $\Gamma_v(0.006 < x < 0.3)$ and
the integral of $g_1^N$ over the same range of $x$ gives a global measurement of the
polarised sea. Indeed, re-ordering Eq.~(11) we obtain
\begin{equation}
  \int_{0.006}^{0.30} \left[ (\Delta {\overline u} + \Delta {\overline d}) + 
  \frac{1}{5} (\Delta s + \Delta {\overline s})\right] {\rm d}x  = 
  -0.02 \pm 0.03~({\rm stat.}) \pm 0.02~({\rm syst.}),
\end{equation}
where the correlation between inclusive and semi-inclusive asymmetries has 
been taken into account in the statistical error. 
This result is compatible with zero but also consistent with the strange
quark contribution of Eq.~(\ref{def_Ds}) and a vanishing contribution from 
the first term.  
It should be kept in mind that moments of sea quarks evaluated at LO
have to be taken with caution 
because their values are small and  thus comparable
to the NLO corrections.

The unmeasured contribution to $\Gamma_v$ for $x > 0.7$ estimated from the LO DNS
parameterisation of Ref.~\cite{dns} is 0.004 at $Q^2 = 10$~(GeV$/c)^2$.
Its upper limit corresponding to the assumption $A_d^{h^+-h^-}$$= 1$ for $x > 0.7$ is 0.007
according to the MRST04 parameterisation.

The unmeasured low $x$ contribution to $\Gamma_v$ is expected to be negligible 
since the integral shows no significant variation when its lower limit is varied 
between 0.006 and 0.02.
We thus estimate the first moment as
\begin{equation}
\Gamma_v (0 < x < 1) = 0.41 \pm 0.07~({\rm stat.}) \pm 0.06~({\rm syst.}).
\end{equation}
The assumption of a fully flavour symmetric sea 
$\Delta {\overline u} = \Delta {\overline d} = \Delta s = \Delta {\overline s}$
obviously leads to $\Gamma_v (0 < x < 1) = a_8$. 
As shown in Fig.~\ref{fig:Adiff_11_release}  (right), 
our experimental value is 
two standard deviations below the value of $a_8  = 0.58 \pm 0.03$ 
derived  from hyperon $\beta$ decays \cite{close}. It has been suggested
that a value of the valence contribution $\Gamma_v$ smaller than  $a_8$
(as expected from the constituent quark models)
could be  a hint that a so far unmeasured part of the nucleon's
spin  resides at $x = 0$ \cite{bass}.

An estimate of the light sea quark contribution to the nucleon spin can be 
obtained by combining the values of $\Gamma_v$ (Eq.~(13)), $\Gamma_1^N$ (Eq.~(1)) and $a_8$
\begin{equation}
\Delta {\overline u} + \Delta {\overline d} = 3 \Gamma_1^N - \frac{1}{2} \Gamma_v + \frac{1}{12} a_8
\end{equation}
and the result is found to be zero (Table~\ref{tab:FirstMom}). 
Possible deviations from the nominal value of $a_8$ due to $SU(3)_f$ symmetry 
violation in hyperon decays are generally assumed to be of the order of 
10\% \cite{Leader} and are included in the systematic error.
The zero value of $\Delta {\overline u} + \Delta {\overline d}$ 
is in contrast with the non-zero value obtained for 
$\Delta s + \Delta {\overline s}$ (Eq.~(\ref{def_Ds})) and  
suggests that $\Delta {\overline u}$ and $\Delta {\overline d}$, 
if different from zero, must be of opposite sign.
Previous estimates by SMC and HERMES, also given in Table~\ref{tab:FirstMom}, 
are compatible with this hypothesis.
The DNS parameterisation  finds a positive $\Delta {\overline u}$
and a negative $\Delta {\overline d}$, about equal in absolute value.
Opposite signs of  $\Delta {\overline u}$ and $\Delta {\overline d}$ are 
predicted in several models, e.g.\ in Ref.~\cite{bbs}
(see also \cite{peng} and references therein).
Forthcoming COMPASS data on a proton target will provide separate determinations
of $\Delta {\overline u}$ and $\Delta {\overline d}$.

In conclusion, we have determined at LO QCD the polarised valence quark 
distribution from the difference asymmetry for oppositely charged hadrons 
in DIS of muons on a polarised isoscalar target. 
Its first moment at $Q^2 = 10$~(GeV$/c)^2$
over the measured range of $x$ (0.006--0.7) is found to be
$ 0.40 \pm 0.07~{\rm (stat.)} \pm 0.06~{\rm (syst.)}$. 
This value disfavours the assumption of a flavour symmetric polarised sea at a 
confidence level of two standard deviations and suggests that $\Delta {\overline u}$ and 
$\Delta {\overline d}$ are of opposite sign.

\section*{Acknowledgements}
We gratefully acknowledge the support of the CERN management and staff and
the skill and effort of the technicians of our collaborating institutes.
Special thanks are due to V.~Anosov and V.~Pesaro for their technical support
during the installation and the running of this experiment.
This work was made possible by the financial support of our funding agencies.


\begin{thebibliography}{}
\bibitem{g1d} COMPASS Collaboration, V.Yu. Alexakhin {\it et al.,}  Phys.\ Lett.\ B {\bf647} (2007) 8.
\bibitem{spectro} COMPASS Collaboration, P. Abbon {\it et al.,}  Nucl.\ Instrum.\ Methods A {\bf577} (2007) 455.
\bibitem{a1d} COMPASS Collaboration, E.S. Ageev {\it et al.,} Phys.\ Lett.\ B {\bf612} (2005) 154.
\bibitem{SMC96} SMC Collaboration, B. Adeva {\it et al.,} Phys.\ Lett.\ B {\bf369} (1996) 93.
\bibitem{SMC98} SMC Collaboration, B. Adeva {\it et al.,} Phys.\ Lett.\ B {\bf420} (1998) 180.
\bibitem{HERMES} HERMES Collaboration, A. Airapetian {\it et al.,} Phys.\ Rev.\ D {\bf71} (2005) 012003.
\bibitem{jetset}T. Sj\H{o}strand, Comp.\ Phys.\ Commun.\ {\bf39} (1986) 347 ; {\bf43} (1987) 367.
\bibitem{aram_1} A. Kotzinian, Phys.\ Lett.\ B {\bf552} (2003) 172.
\bibitem{aram_2} A. Kotzinian, Eur.\ Phys.\ J. C {\bf44} (2003) 211.
\bibitem{Frankfurt} L.L. Frankfurt {\it et al.,} Phys.\ Lett.\ B {\bf230} (1989) 141.
\bibitem{Christova} E. Christova, E. Leader, Nucl.\ Phys.\ B {\bf607} (2001) 369.
\bibitem{SSI} A.N. Sissakian, O.Yu. Shevchenko, O.N. Ivanov, Phys.\ Rev.\ D {\bf73} (2006) 094026.
\bibitem{poldis} A. Bravar, K. Kurek, R. Windmolders, Comput.\ Phys.\ Commun.\ {\bf105} (1997) 42.
\bibitem{mrst} A.D. Martin, W.J. Stirling, R.S. Thorne, Phys.Lett.\ B {\bf636} (2006) 259.
\bibitem{abe} E143 Collaboration, K. Abe {\it et al.,} Phys.\ Lett.\ B {\bf452} (1999) 194.
\bibitem{mch} R. Machleidt {\it et al.,} Phys.\ Rep.\ {\bf149} (1987) 1.
\bibitem{dns} D. de Florian, G.A. Navarro, R. Sassot, Phys.\ Rev.\ D {\bf71} (2005) 094018.
\bibitem{aac} AAC collaboration, M. Hirai, S. Kumano, N. Saito, Phys.\ Rev.\ D {\bf69}
(2004) 054021.
\bibitem{kretzer} S. Kretzer, Phys.\ Rev.\ D {\bf62} (2000) 054001.
\bibitem{kkp} B.A. Kniehl, G. Kramer, B. Potter, Nucl.\ Phys.\ B {\bf582} (2000) 514.
\bibitem{close} F.E. Close, R.G. Roberts, Phys.\ Lett.\ B {\bf316} (1993) 165.
\bibitem{bass} S. D. Bass, Eur.\ Phys.\ J. A{\bf5} (1999) 17;
 S.D. Bass, Rev.\ Mod.\ Phys.\ {\bf77} (2005) 1257.
\bibitem{Leader} E. Leader, D. Stamenov, Phys.\ Rev.\ D {\bf67} (2003) 037503.
\bibitem{bbs} C. Bourrely, F. Buccella, J. Soffer, Eur.\ Phys.\ J. C {\bf41} (2005) 327.
\bibitem{peng} J.C. Peng, 
               Eur.\ Phys.\ J. A{\bf18} (2003) 395.


\end{thebibliography}
\end{document}